\documentclass{article}

\usepackage{arxiv}

\usepackage[utf8]{inputenc} 
\usepackage[T1]{fontenc}    
\usepackage{hyperref}       
\usepackage{url}            
\usepackage{booktabs}       
\usepackage{amsfonts}       
\usepackage{nicefrac}       
\usepackage{microtype}      
\usepackage{lipsum}
\usepackage{graphicx}
\usepackage[authoryear]{natbib}
\usepackage{xcolor}
\usepackage[acronym]{glossaries}
\usepackage{listings}
\usepackage{subfig}
\usepackage[title]{appendix}

\begin{document}
\title{Flo: A data-driven limited-area storm surge model}   
\author{
Nils Melsom Kristensen\\
Norwegian Meteorological Institute\\
Oslo, Norway\\
\texttt{nilsmk@met.no}\\
\And
Mateusz Matuszak\\
Norwegian Meteorological Institute\\
Oslo, Norway\\
\And
Paulina Tedesco\\
Norwegian Meteorological Institute\\
Oslo, Norway\\
\And
Ina Kristine Berentsen Kullmann\\
Norwegian Meteorological Institute\\
Oslo, Norway\\
\And
Johannes Röhrs\\
Norwegian Meteorological Institute\\
Oslo, Norway\\
}

\maketitle
\begin{abstract}
We present \textit{Flo}, a data-driven storm surge model, covering the North Sea, Norwegian Sea and Barents Sea. The model is built using the Anemoi framework for creating machine learning weather forecasting systems, developed by the European Centre for Medium-Range Weather Forecasts and partners. The model is based on a graph neural network, and is capable of simulating water level due to atmospheric effects (wind stress and inverse barometer effect, i.e. the non-tidally induced part of the total water level; the residual water level) at a horizontal resolution of 4 km and a temporal resolution of 1 hour with a quality comparable to the numerical model on which it was trained. The model was trained using a dataset consisting of 43 years of atmospheric data from the 3-km Norwegian Reanalysis hindcast for mean sea level pressure and winds, and the NORA-Surge hindcast for water level. Evaluation was done by comparing results from hindcast runs of the Flo model against independent observations of more than 90 water level gauges along the European coast, and against the NORA-Surge hindcast. The evaluation shows that Flo produces hindcasts with accuracy similar to the NORA-Surge hindcast, and it is shown that the model can resolve key physical processes. As the NORA-Surge hindcast used for training does not include data assimilation, Flo is not expected to systematically outperform the numerical model when evaluated against observations. Nevertheless, the present work represents an important step towards complementing traditional physics-based storm surge modelling with machine learning approaches and the framework establishes a strong foundation for future developments, particularly for training storm surge models that offer more flexibility for incorporating observations and other additional data sources.
\end{abstract}


\maketitle

\section{Introduction}\label{sec:intro}
Neural networks (NN) have been increasingly applied to storm surge prediction and post-processing, particularly for the correction of systematic model biases (e.g., \cite{tedesco23, Zhu2025PINN,Hermans2025Europe,Naeini2025HDNN}). In addition, recent advances utilizing machine learning (ML) techniques and data-driven models (DDM) within the field of meteorology have resulted in a new generation of weather forecasting models based on graph neural networks (GNNs) that are efficient and lightweight to run, and in many cases surpass the forecast quality of their numerical weather prediction (NWP) model counterparts \citep{keisler2022,bi2023,chen2023,kurth2023,Ben_Bouall_gue_2024,lang2024,nipen2025}. The latter is a direct result of the fact that many of these DDMs are trained on reanalyses as training datasets. A reanalysis is largely influenced by the assimilation of observations, and hence, a DDM trained on such data, is not limited by the discretizations of the governing equations of motion, which are the foundation for most NWP models. 

Building on this line of research, we develop a high resolution limited-area model (LAM) for predicting water levels due to atmospheric effects (wind stress and inverse barometer effect, i.e., the non-tidally induced part of the total water level; the residual water level) based on a GNN using the Anemoi framework for creating machine learning weather forecasting systems \citep{lang2024}. 

The GNN architecture is well suited for applications like storm surge forecasting due to its flexibility and ability to build native spherical grids, with the possibility of using non-uniform grids, and multiple messaging passing levels to enable connections across several distances.

Residual water level is the atmospheric contribution to the sea surface height (SSH) due to the inverse barometer effect \citep{wunsch1997} and the horizontal displacement of water masses by wind stress acting on the water surface \citep{kristensen2024}. Together with the periodic and highly predictable astronomical tides, the atmospheric effect is the main contribution to the total water level observed along coastlines. Due to its direct dependence on the actual atmospheric conditions, it has to be forecasted on a day-to-day basis using operational storm surge models (e.g \cite{zijl:etal:2013, kristensen22}). Extreme water levels that can pose a risk to life and property are most often caused by storm surge, often in combination with high tides. Storm surge prediction is therefore aimed at skillfully describing extreme values. However, areas along the coastal zone have different tolerance thresholds for water level, and what is considered an extreme in terms of impact on infrastructure varies geographically \citep{fang2014}. Storm surge is a barotropic process, i.e., depth-integrated shallow water dynamics apply, acting on the entire water column. Storm surge can therefore be accurately simulated using barotropic two-dimensional ocean models. 
Due to the good quality of the present-day weather models, and as a consequence of their critical dependence on atmospheric conditions, storm surge is highly predictable at synoptic time scales. 

In addition to operational forecasting, storm surge models are also used to provide datasets of historical storm surge, i.e., hindcasts. Observation datasets consisting of data from water level gauges are by many considered as the preferred product for knowledge about historical storm surge. However, these datasets often have poor coverage both geographically and temporally, therefore models fulfill a crucial role to provide spatio-temporal coverage.

Hindcasts are used for various purposes, e.g., calculating extreme and return values \citep{kristensen2024} and defining safety zones for infrastructure located in the coastal zone. Recently, hindcasts and reanalyses are increasingly being utilized as training datasets for DDMs (e.g., \cite{graphcast2023, lang2024, nipen2025}). The present work makes use of the NORA-Surge hindcast \citep{kristensen2024} as training data in the training of a DDM for storm surge modeling, the \textit{Flo} model. The model area, as shown in Figure \ref{fig:model-domain}, covers roughly the area between the western entrance to the English Channel, Novaya Zemlya and the east coast of Greenland, including the North Sea, Norwegian Sea, and the Barents Sea. This is an area where intense low-pressure systems regularly pass trough. Combined with the fact that the coastline around the North Sea in general is a flat coast, where large areas such as e.g. the Netherlands are located very close to the sea level height, the area is prone to inundation due to storm surge. Since many people live, and much of the infrastructure is located at, or close to, the coast, this highlight the importance of accurate and efficient storm surge forecasting in this area.

Previous efforts to utilize ML for predicting storm surge in our study area include e.g. \cite{tiggeloven_exploring_2021} where  they used an Artificial Neural Network (ANN), a Long Term Short Term Neural Network (LSTM), a Convolutional Neural Network (CNN), and a Convolutional LSTM (ConvLSTM, a combination of CNN and LSTM) to predict storm surge globally. More recently \cite{Hermans2025Europe} utilized LSTM  and ConvLSTM models to compute extremes at 9 locations in Europe with similar skill as the Global Tide and Surge Model (GTSM) hydrodynamical model. A common property for most previous work on this subject, including \cite{tiggeloven_exploring_2021} and \cite{Hermans2025Europe}, is that predictions are only available at the a priori specified tide-gauge locations used in the training of the model. Whereas our approach, utilizing the GNN framework, enable us to produce consistent two-dimensional fields of water level for the entire model domain at high resolution, with the ability to provide a prediction of the storm surge at any arbitrary location.

In the current study, we train the Flo model on hindcast data, and evaluate it by running inference to produce a water level hindcast, forced by an atmospheric hindcast. Alternatively, the model can also be used to produce water level forecasts, by forcing it with atmospheric forecasts, even if not demonstrated here.

The article is structured as follows. Section \ref{sec:method} describes the model setup, including the training data and the model architecture of the Anemoi GNN model, and the training and inference of the model. Further, Section \ref{sec:eval} contains details about the observation dataset used for the model evaluation, and the evaluation of the model performance compared against both the NORA-Surge hindcast and observations. Section \ref{sec:summary} provides a summary and some concluding remarks.

\section{Method}\label{sec:method}
\subsection{Training data}\label{sec:traindata}

The Flo model is built using the Anemoi framework \citep{lang2024}, which is a complete software suite, written mainly in Python, that handles all the steps of creating machine learning weather and ocean models in a user-friendly way. It is a LAM configuration in Anemoi, which means it uses training data for a regional domain covering the North Sea, Norwegian Sea and Barents Sea (see Figure \ref{fig:model-domain}). 
The data used for training consists of two datasets: 1) the NORA3 atmospheric hindcast \citep{haakenstad21nora3} and 2) the NORA-Surge storm surge hindcast \citep{kristensen2024}. 
Variables used in the training, see Table \ref{tab:trainvars}, are the SSH from the NORA-Surge hindcast and, as forcing variables, 10-meter winds as x- and y-components (U10m and V10m) and mean sea level pressure (MSLP) are taken from NORA3. All variables are interpolated onto a common grid with a horizontal resolution of 4 km. The temporal resolution of the training data is 1 hour. We note that the NORA-Surge hindcast was forced by wind stress calculated from NORA3 winds based on the Charnock relation \citep{charnock:1955}, while in the present work, for convenience, we have opted to use the wind directly in the training.
Other forcing variables include bathymetry, sea-land mask, and the Coriolis parameter. In addition, other standard ML-forcings include insolation, and the sine and cosine of latitude, longitude, local time and Julian day. 

A total of 43 years of data are available from the NORA3 and NORA-Surge hindcasts.
The model training uses data from the 16-year time period 1990-2005 (inclusively), while the validation dataset, used for training validation metrics, consists of the four years 2006-2009.
The remaining 13 years, 2010-2022, were reserved for out-of-sample evaluation of the model and case studies. The selected periods, and their lengths, are in part chosen to match the time periods with large amounts of observation data to compare with, and to enable us to complete the training within a reasonable time on the available compute resources.

\begin{figure}
  \centering
  \subfloat[]{\includegraphics[width=0.5\textwidth]{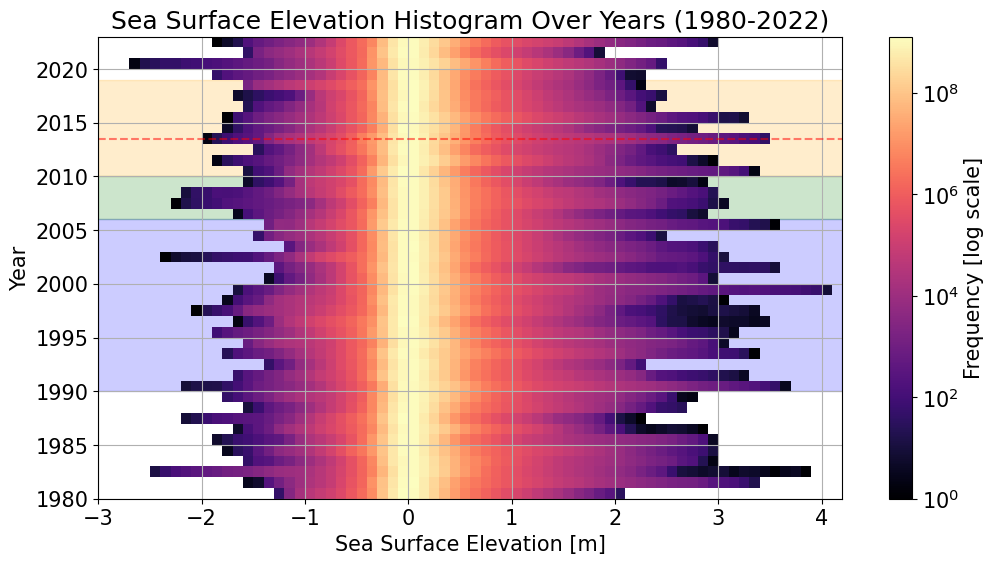}\label{fig:train_hist_2d}}
  \subfloat[]{\includegraphics[width=0.5\textwidth]{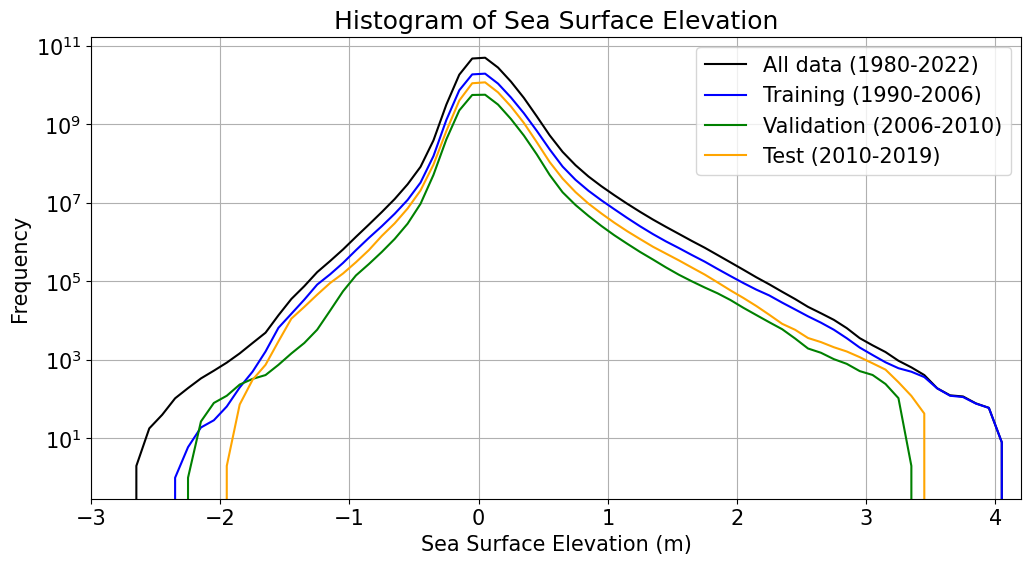}\label{fig:train_hist}}
  \caption{Histograms showing the distribution of residual water level in the entire NORA-Surge training dataset, for all times and all grid points. Panel (a) shows a 2D year-by-year comparison of residual water level distribution (values on the color scale are logarithmic), whereas panel (b) show combined distribution for the relevant training, validation and test periods, also on a log scale. The blue, green and yellow background colors in panel (a) depict the training, validation and test periods, respectively. In addition, the horizontal dashed red line marks the year 2013 that contains the storm Xaver.}
  \label{fig:train_stats}
\end{figure}

Figure \ref{fig:train_hist_2d} shows the statistical distribution of residual water level for each year of the NORA-Surge dataset (all grid points), with the training, validation and evaluation periods marked with a blue, green and yellow background, respectively. In Figure \ref{fig:train_hist} we show the summarized distribution of residual water level for the training, validation and evaluation periods. The dataset contains, for each time step, a total number of $593\,920$ grid points, whereas $436\,047$ of them (approximately $73\%$) are ocean points. 
It is clearly evident in Figure \ref{fig:train_stats} that the residual water level is very close to zero most of the time. However, the socio-economic impact of storm surge, including consequences to life and property, becomes increasingly important as the amplitude of the residual water level increases, i.e., it is a case of extreme value prediction. Figure \ref{fig:train_stats} shows that even in a 43-year long dataset, extreme values are very rare.

Usually, LAM-models implement lateral boundary forcings using a lower-resolution model surrounding the regional domain (e.g., see \cite{oskarsson2023,adamov2025,wijnands2025} which uses coarser global models as boundary forcing datasets). In an operational setup, a data-driven LAM model requires boundary forcings at each forecast time step, which may be provided by an external numerical model. In the current work, we have taken a simpler approach, where we have reduced the size of the prediction domain by five grid points, and used the remaining cells as boundary forcings. This means that the state of boundary domain is provided by the same dataset as the interior domain, and that the resolution is the same. This setup may be extended to facilitate an operational setup, as further discussed in \ref{sec:summary}.

\subsection{Model architecture} \label{sec:modelarch}
\begin{figure}
    \centering
    \includegraphics[width=0.5\linewidth]{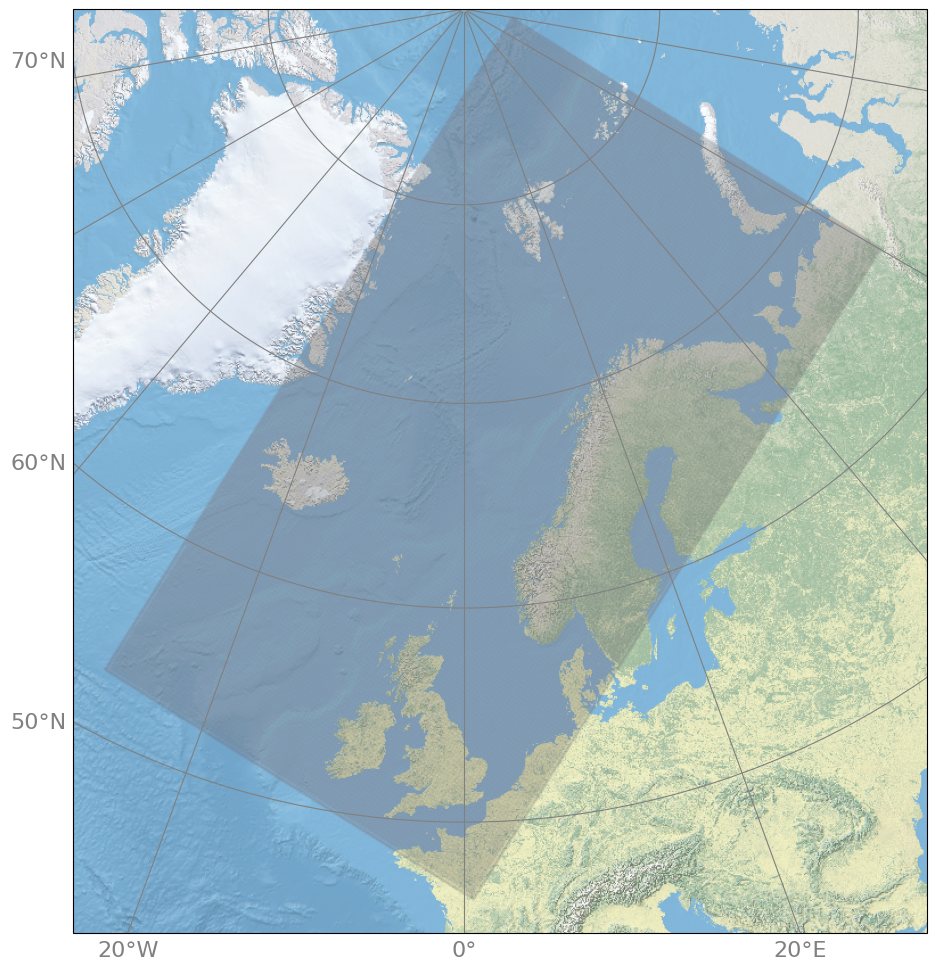}
    \caption{The storm surge model domain covering the North Sea, the Norwegian Sea and the Barents Sea is shown by the shaded area.}
    \label{fig:model-domain}
\end{figure}

\begin{figure}
  \centering
  \subfloat[Encoder]{\includegraphics[width=0.5\textwidth]{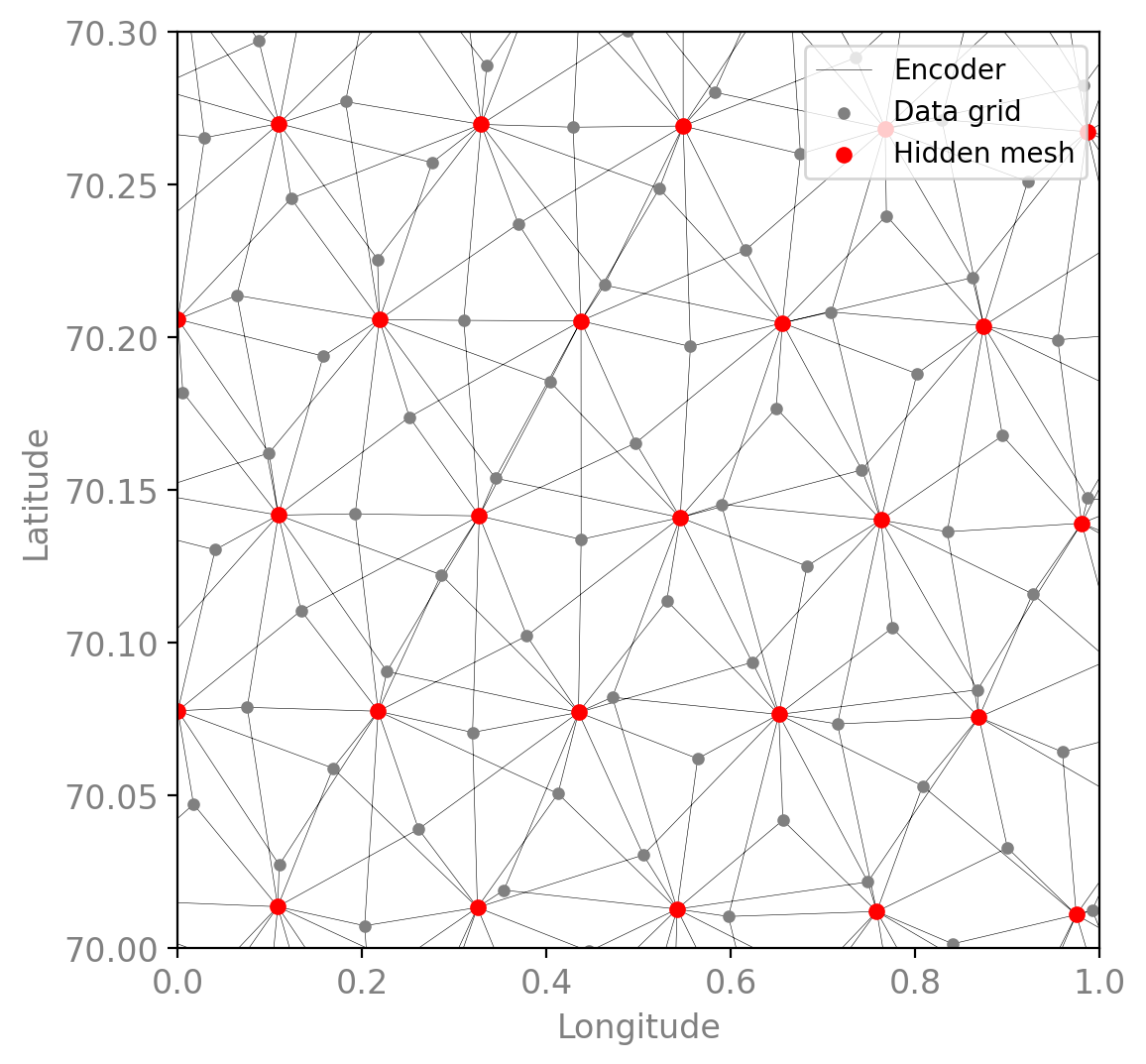}\label{fig:encoder}}
  \subfloat[Decoder]{\includegraphics[width=0.5\textwidth]{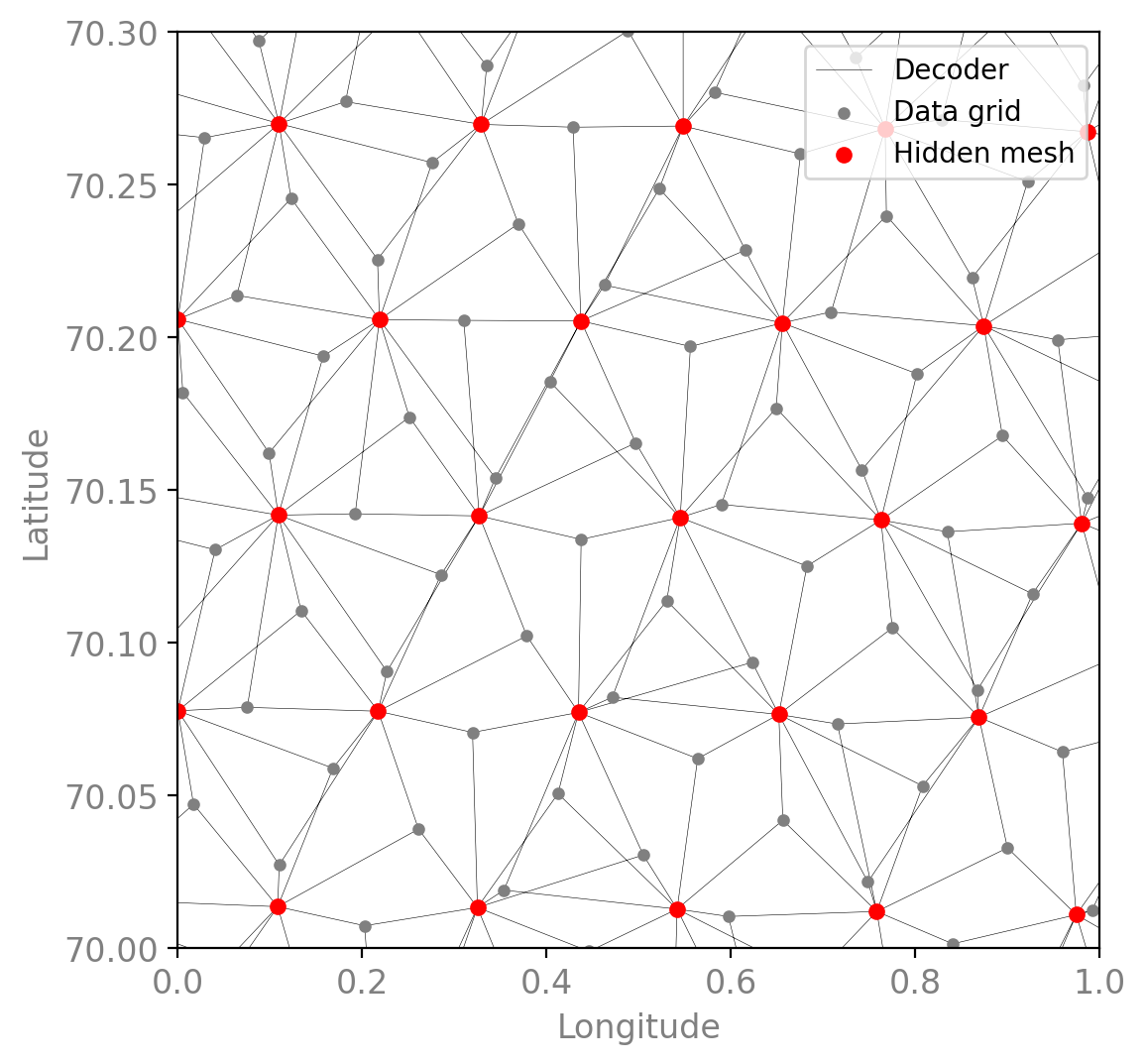}\label{fig:decoder}}
  \caption{Example of how the mapping of the physical model grid (data grid) space is mapped into the hidden mesh via the encoder in panel (a) and back into data grid via decoder in panel (b).}
  \label{fig:enc-dec}
\end{figure}

\begin{figure}
  \centering
  \subfloat[Processor outer]{\includegraphics[width=0.5\textwidth]{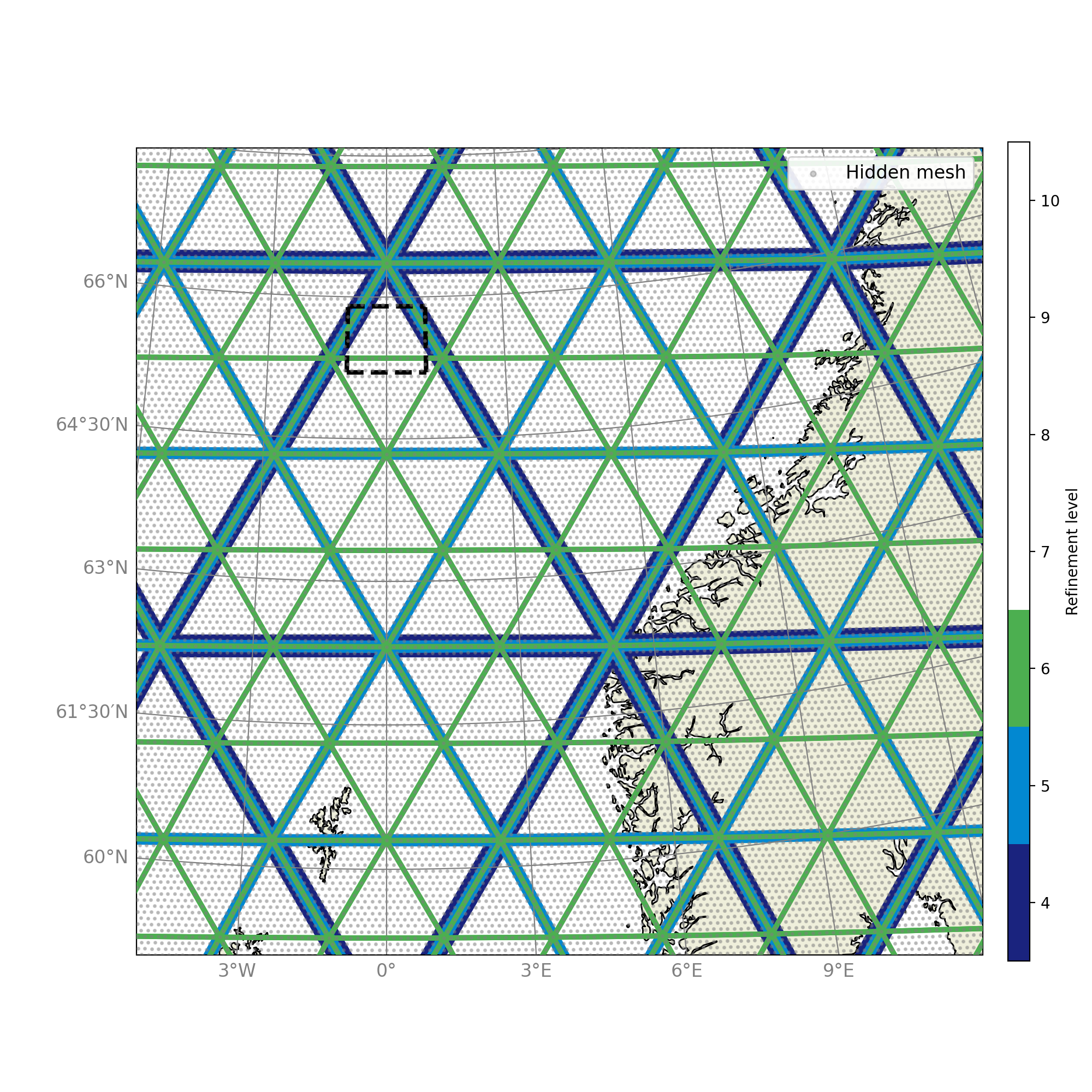}\label{fig:processor_outer}}
  \subfloat[Processor inner]{\includegraphics[width=0.5\textwidth]{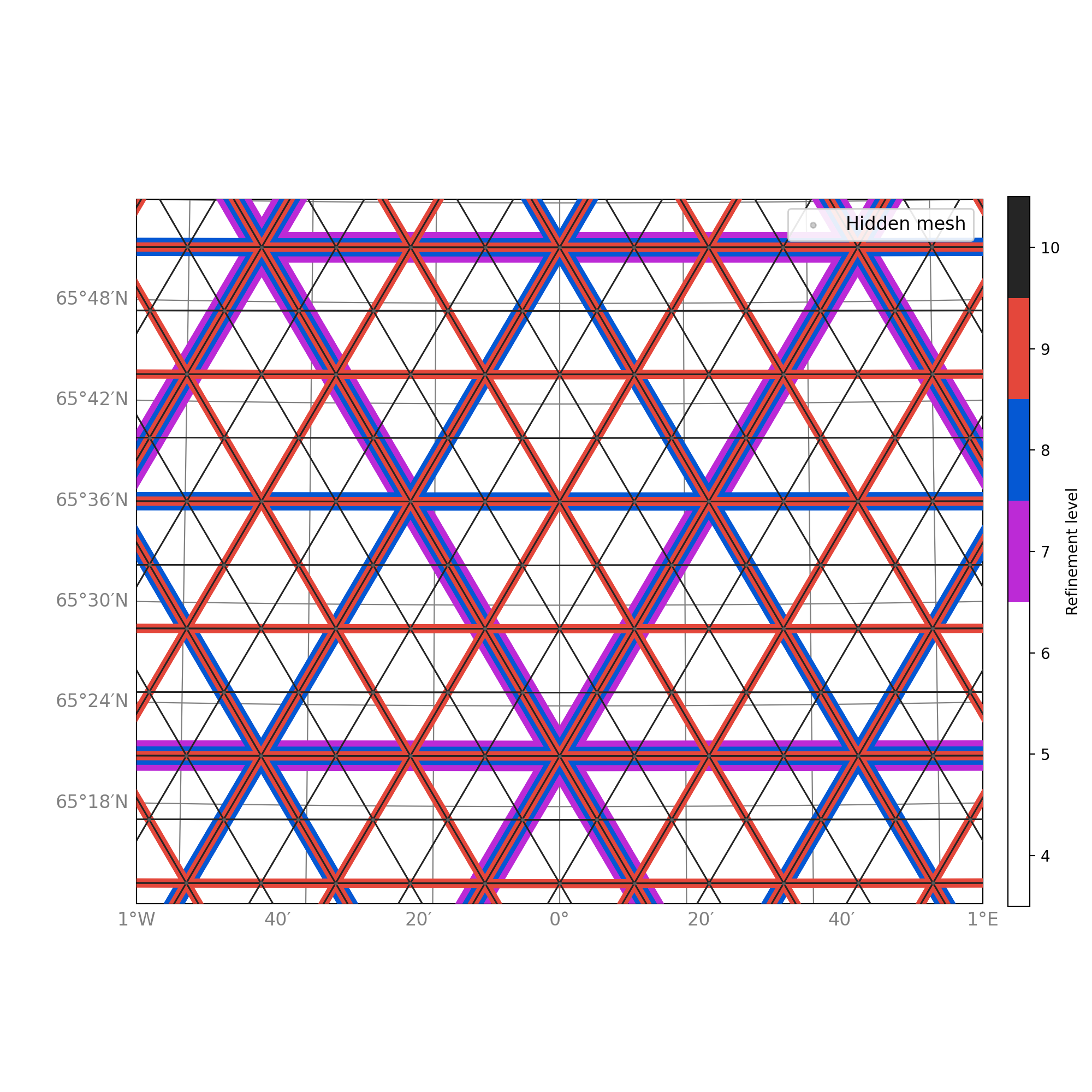}\label{fig:processor_inner}}
  \caption{Visualization of how the processor connects the nodes of the hidden mesh via edges at different refinement levels. Panel (a) shows the edges between nodes at refinement level 4-6, where lower number equals longer edges. Panel (b) shows the edges and connections between the nodes at refinement levels 7-10 for the area outlined by the square of dotted lines in panel (a).}
  \label{fig:processor}
\end{figure}

Flo is based on an encoder-processor-decoder architecture, similar to the works of \cite{graphcast2023,nipen2025,lang2024}.
The encoder transforms the input state on the NORA-Surge grid (see Figure \ref{fig:model-domain}), which consists of surface variables and weather forcings relevant to residual water level prediction (see Table \ref{tab:trainvars}), into a lower-dimensional latent space (see Figure \ref{fig:encoder}). Within this latent space, referred to as the hidden mesh (see Figure \ref{fig:processor}), the processor advances the ocean state forward in time. Finally, the decoder maps the predicted latent state back to the original data grid (as shown in Figure \ref{fig:decoder}), providing the desired residual water level variables on the input grid.
Within Anemoi, we apply the graph transformer option for the model with 16 attention heads, as described in \citet{lang2024}, which allows to propagate information over longer distances.

The graph is a central piece of the GNN architecture, consisting of nodes in data and latent space that are connected by edges. The data nodes are created directly from the grid points of the NORA3 and NORA-Surge training data. Following \cite{lang2024} and \cite{graphcast2023}, the processor mesh nodes are represented by the vertices derived from iterative refinements of an icosahedron around the globe, see Figure \ref{fig:processor}. At base refinement $N=0$ we obtain an icosahedron with $12$ vertices, $30$ edges and $20$ faces. Each increase in refinement level divides the existing faces into four smaller ones, increasing the number of vertices, and thus the number of processor mesh nodes. The number of vertices $V_N$, or mesh nodes, at refinement level $N$ can be expressed as:

\begin{equation}
    V_{N} = 2+10\cdot 4^{N}.
\end{equation}

Our mesh is constructed at refinement level $10$, yielding $10\,485\,762$ global processor mesh nodes. However, as our model uses a LAM setup, we remove any processor mesh nodes outside the region of interest, yielding $211\,525$ processor mesh nodes, roughly one third as many processor nodes as grid nodes.

The encoder maps information from data nodes to hidden nodes through the edges, see Figure \ref{fig:encoder}. 
To create the edges of the encoder, we evaluate the distance between two neighboring processor nodes. This distance is the largest permitted encoder edge length. Encoder edges are created to connect each processor mesh node to all data grid nodes within this distance. Thus, a single processor mesh node will obtain information from multiple data grid nodes, and data grid nodes send information to multiple processor mesh nodes. To reduce the complexity of the graph, the maximum number of data nodes that a mesh node may be connected to is $64$. Thus, if the number of edges reaches $64$ during edge creation, the algorithm will stop, which may leave unconnected nodes called orphans. At refinement level $10$, we do not observe such unconnected nodes. 

Similarly to the creation of the processor mesh nodes described above, the processor edges are directly obtained from the edges of the iteratively refined icosahedron (e.g., Figure~\ref{fig:processor}). For a global model with refinement $N=0$, each processor node will communicate with the five neighboring processor nodes. Each increase in refinement level $N$ creates new nodes that communicate with their neighbors. The length $L$ of a processor edge at each refinement $N$ can be described as:
\begin{equation}
    L_N = \frac{L_{N-1}}{2},
\end{equation}
i.e., for each increase in refinement level, the edge lengths are halved. Processor edges from all iteratively constructed refinement levels are accumulated.

For our LAM domain, we remove any edges and nodes outside of the model domain (see Figure~\ref{fig:model-domain}). This effectively removes all edges at $N=0$ as these are too long.  
Even if storm surge is a fast-moving phenomenon that effectively travels the length of the model domain in less than 24 hours, the few scattered long edges on refinement level 1-3 only provide minimal coverage due to the orientation of the polar stereographic domain. 
Thus, we decided to only keep processor edges at refinement levels $N\geq4$, which further reduces the complexity of our graph. 

The decoder maps information from the processor mesh in latent space back to the data grid. Our decoder connects each data grid node to the three closest processor mesh nodes by using the k-nearest neighbors algorithm (KNN). Our LAM setup uses a boundary region to provide the model with lateral boundary conditions. Since we do not want to predict the ocean state at the domain edge, we construct our decoder such that no mesh nodes are connected to data nodes in the boundary region. 

\subsection{Training} \label{sec:trainconfig}
We have performed two trainings during the development of the Flo model. One is a training based on 16 years (1990-2005) of data, hereafter referred to as the Flo model, while the other was based on the 4 years 1990-1993. Even if the NORA-Surge hindcast covers a longer time period, we have selected the years starting from 1990 because more observations are available from 1990 and onward. The two training runs enable us to evaluate if we should prioritize a longer training period containing more data, or if running a shorter period with less data for more epochs provides better results given the same amount of GPU resources. Both training periods could be considered to be short, but based on the distribution of residual water level for the years that make up the training dataset, as shown in Figure \ref{fig:train_stats}, it is clear that both of the selected training periods has a distribution of residual water level that in a good way represent the overall distribution. In addition, we suspect that the physical processes that drive the residual water level (i.e., the inverse barometer effect and the advection due to wind stress) are easily identified and learned during model training without explicitly including physical processes in the loss function (as is the case with so called Physics-informed neural networks), and do not require a very long training period with many extremes in order to be able to simulate extremes. By learning the physics that govern the training data instead of the statistical distribution itself, we expect Flo to be able to predict extremes far beyond what is present in its training data. This can be done as long as the governing physics that produce the extremes is adequately covered by the data in the training, i.e. that e.g. there are no non-linear effects that are not properly described.

The model was trained on a single Nvidia H200 GPU using Anemoi-training version 0.6.4 on local infrastructure at MET Norway. We used a learning rate of $6.25\times10^{-4}$, and the model was run for $10^{5}$ steps with a batch size of $4$. The number of training steps was chosen based on what was considered an acceptable computation time given the available time and resources. The warmup period was set to $1\,000$ steps, and the minimum learning rate of the cosine learning rate schedule was $3.0\times10^{-7}$ and the number of channels was $256$. The same setup was used for both the trainings, and results in $\sim3$ and $\sim12$ epochs for the long and short training periods, respectively. Each of the two trainings consumed $142$ GPU hours.


\begin{figure}
  \centering
  \includegraphics[width=0.75\linewidth]{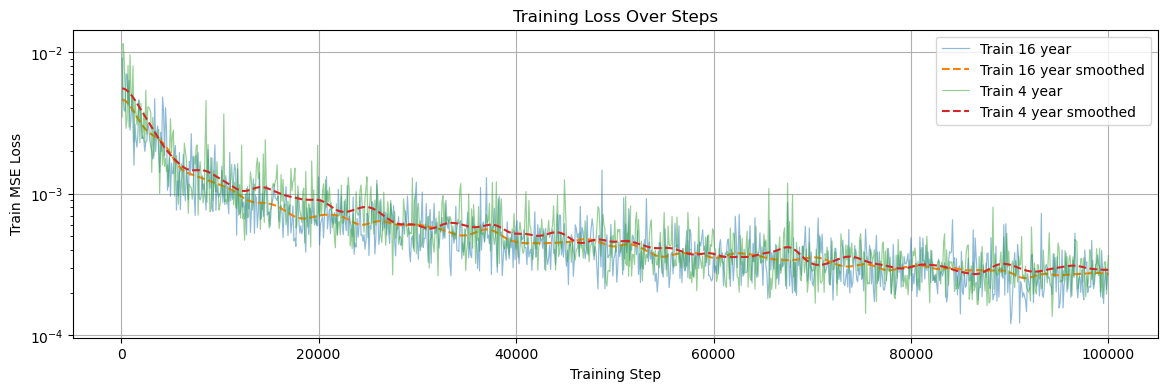}
  \caption{The training loss for every 100 training steps as a function of steps, as calculated by Anemoi-training. We show the training loss for both of the training runs, together with a smoothed line for each of the two.}
  \label{fig:trainloss}
\end{figure}

The training MSE loss, as calculated by the Anemoi framework during training, for the two runs are displayed in Figure \ref{fig:trainloss} on a log scale. The loss starts at around $5.0\times10^{-3}$ for both trainings, and quickly drops below $10^{-3}$ during the first 1500-2000 steps. It is clear how the loss curves still have a negative trend even towards the end of the training period, suggesting that the training could benefit from running for more steps.

\subsection{Inference}\label{sec:infer}
The evaluation of the Flo model, presented in Section \ref{sec:eval}, is based on inference of the model as a series of hindcast runs. Firstly, we perform a continuous run of the DDM, initialized from NORA-Surge at 2010-01-02 00 UTC and run for 78840 hours until 2018-12-31 00 UTC. The inference was forced by NORA3 10-meter winds and MSLP, and the SSH boundary data were taken from the NORA-Surge hindcast. This was done to streamline the process of performing the inference runs, and to ensure consistency with NORA-Surge. However, this way of using the host model as boundary conditions during inference is not necessarily possible in e.g. an operational system, and hence this information will have to be supplied by an outer, preferably global, ocean model or through a parameterization of the SSH using e.g. the IB effect as described in \cite{kristensen2024}.
The inference runs were done on the same H200 GPU as the training.
We performed the hindcast runs twice, based on the training checkpoints from both the 4 and 16 years long training runs after $10^{5}$ steps.

In addition, we have conducted a series of shorter inferences of 240 hours length. Inference was run for the checkpoints after 25, 50, 75 and $100\times10^{3}$ steps for both training runs. These were re-initialized from NORA-Surge every 10 days during the year 2013, resulting in 36 runs for each of the checkpoints, a total of 288 runs. Atmospheric and boundary forcing were the same as for the longer hindcast runs. This was done to explore at which rate the two training runs converge towards NORA-Surge.

\subsection{Observation dataset}\label{sec:obsdata}
\begin{figure}
    \centering
    \includegraphics[width=0.25\linewidth]{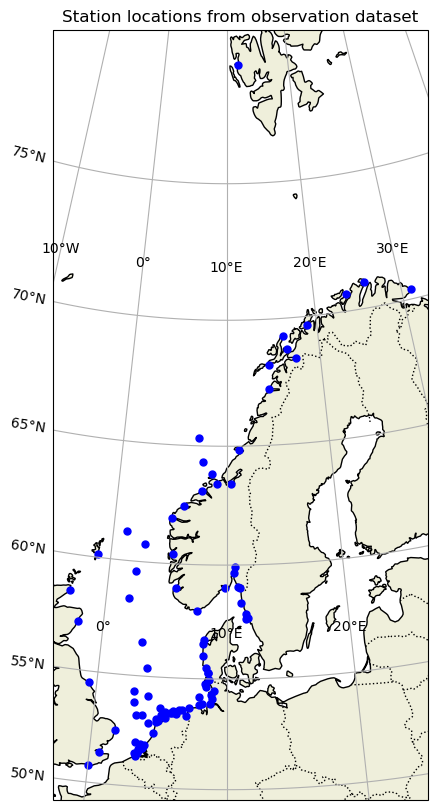}
    \caption{Positions of the observation locations used in the analysis.}
    \label{fig:obsloc}
\end{figure}

The observation dataset used to evaluate the quality of the Flo model consists of a collection of water level observations from 98 tidal gauge stations around the North Sea, Norwegian Sea and Barents Sea (see Figure \ref{fig:obsloc}). The same dataset was quality controlled and used for validation by \cite{tedesco23} and \cite{kristensen2024}.
The observation dataset is de-tided using harmonic analysis, as described by \cite{kristensen2024}, and the following analyses focus purely on the residual water level.

\section{Results and Discussion}\label{sec:eval}

The quality of the Flo model is evaluated for the time period 2010-2019, which has not been seen by the model during training. We present here a direct comparison with the results from the same numerical model on which Flo was trained, and a comparison with water level observations.
We note that we do not expect superior performance of the DDM compared to the hindcast, contrary to many DDMs for atmospheric prediction that perform better than their NWP model counterparts due to training on reanalysis (e.g. \cite{graphcast2023,lang2024,Ben_Bouall_gue_2024,nipen2025}). We do not expect Flo to perform better than the physical model since the training is not based on a reanalysis correcting for model biases through observations. Instead, the  DDM is used to emulate the physics of the numerical model directly. Hence, a "perfect" result would be that the DDM produces exactly the same fields as those obtained by running the numerical model, i.e., the NORA-Surge hindcast.

\subsection{Comparison against NORA-Surge}\label{sec:compare_mod}
\begin{figure}
  \centering
  \subfloat[]{\includegraphics[width=0.75\textwidth]{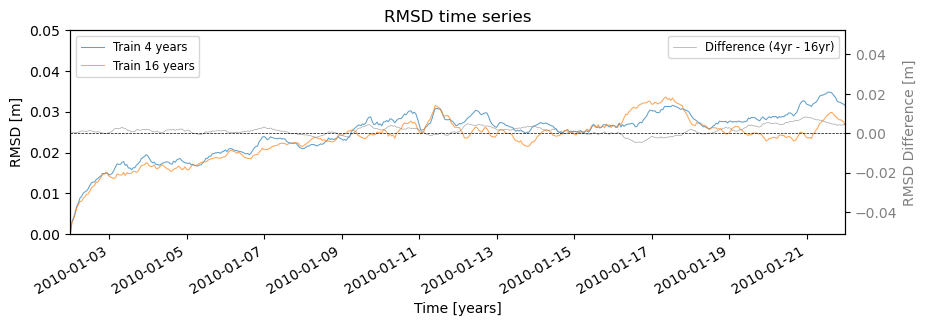}\label{fig:rmsd_longrun_a}}\hfill
  \subfloat[]{\includegraphics[width=0.75\textwidth]{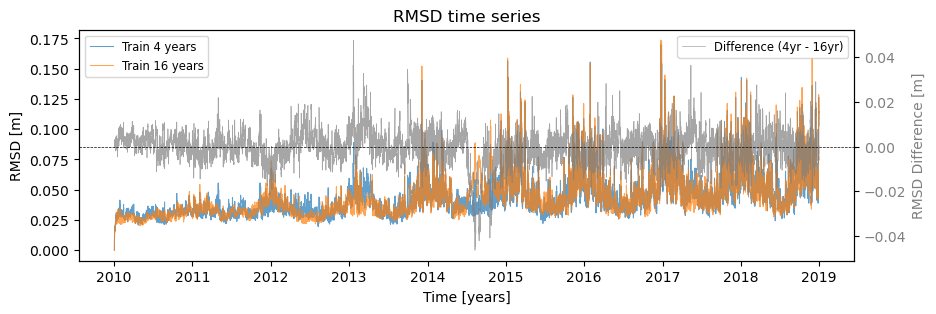}\label{fig:rmsd_longrun_b}}
  \caption{Time series of RMSD between Flo and NORA-Surge at every timestep during the inference runs (2010-2018). Panel (a) shows the first 480 hours of the inference run, while panel (b) shows the entire time period.}
  \label{fig:rmsd_longrun}
\end{figure}

To evaluate the difference between Flo and NORA-Surge, we compute the root-mean-squared-differences (RMSD) between the two for every grid point for all time steps: Figure \ref{fig:rmsd_longrun} shows that the RMSD increase during the first 24-48 hours, and stabilize between 2 and 3 cm for the remainder of the initial period (Figure \ref{fig:rmsd_longrun_a}). This could be an effect of smoothing, which is a known behavior in many ML-models \citep[e.g.]{graphcast2023}. A careful examination of the RMSD evolution in Figure \ref{fig:rmsd_longrun_b} reveal a few key observations: Firstly, the RMSD exhibits seasonality and secondly,  peaks in RMSD increase for lead time in the inference. In addition, there is a slight positive trend in the RMSD. The comparison between the two trainings based on 4 and 16 years of data suggests that the longer training reveals slightly lower RMSD on average, but generally differs little.

\begin{figure}
    \centering
    \includegraphics[width=0.5\linewidth]{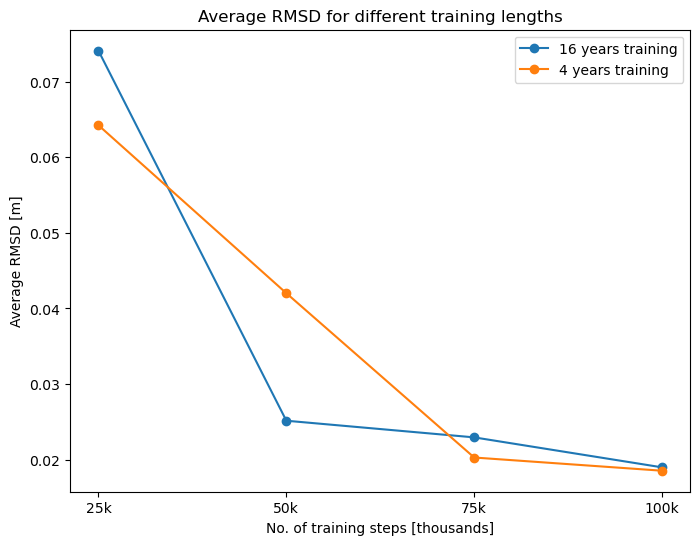}
    \caption{RMSD between inference of the Flo model based on the two training runs (4 and 16 years) at different steps of the training runs, and NORA-Surge. The RMSD, in meters, is calculated as an average over all grid points at every hour for the entire 240-hour inference run for each of the 36 inference runs performed for each of the checkpoints for both trainings (as described in Section \ref{sec:infer}).}
    \label{fig:rmsd240h2013}
\end{figure}

In addition, as mentioned in Section \ref{sec:trainconfig}, we explore how the two different lengths of the training dataset and the number of training steps affect the model performance. Figure \ref{fig:rmsd240h2013} shows a comparison of the average RMSD between inference of the Flo model and NORA-Surge based on the two training runs at four different checkpoints as described previously in Section \ref{sec:infer}. This suggests that a similar quality for the DDM is obtained, as long as the runs are sufficiently long. The test in Fig. \ref{fig:rmsd240h2013} indicates that the DDM based on the long training period converges faster, but we emphasize that the rate of convergence can depend on the time period and cannot be generalized.

\subsection{Evaluating barotropic wave propagation}

\begin{figure}
    \centering
    \includegraphics[width=0.5\linewidth]{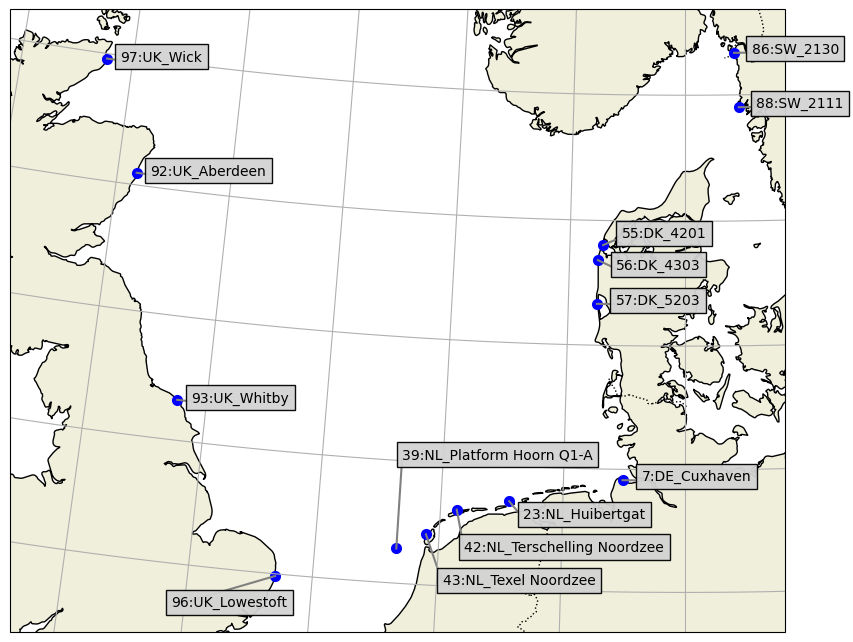}
    \caption{Map showing the selected stations used in the analysis in Figure \ref{fig:hovmoeller} and \ref{fig:along_wave}.}
    \label{fig:map_selected}
\end{figure}

\begin{figure}
    \centering
    \subfloat[]{\includegraphics[width=0.75\linewidth]{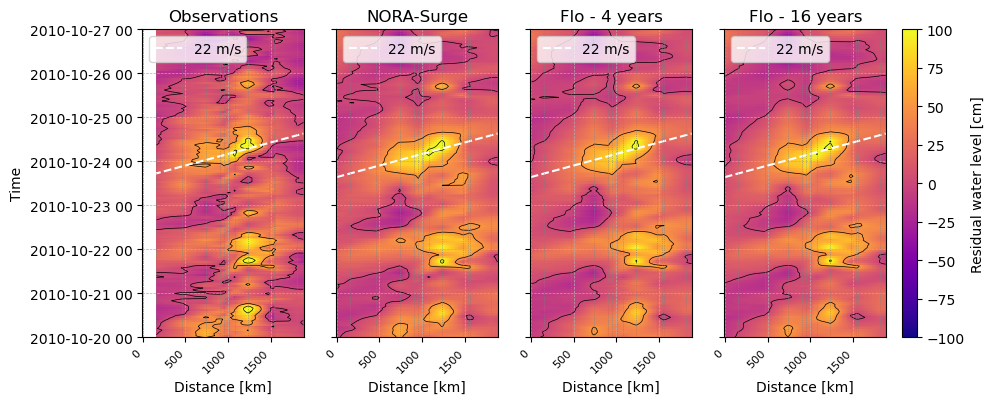}}\hfill
    \subfloat[Xaver]{\includegraphics[width=0.75\linewidth]{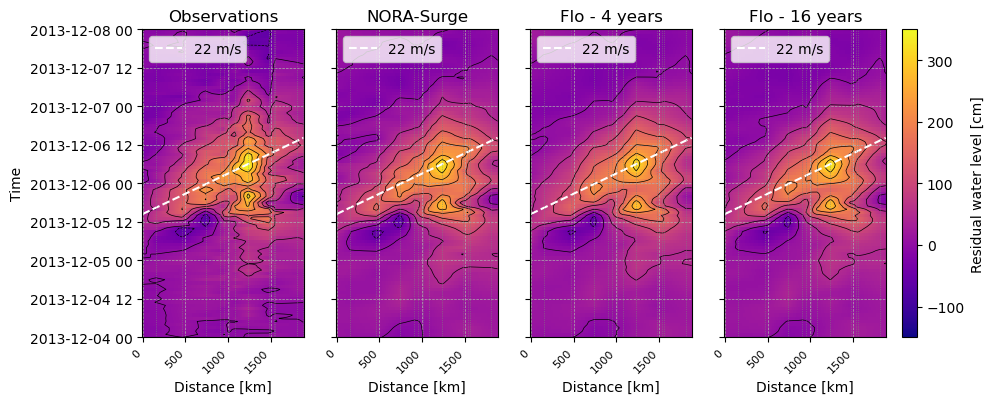}}
    \caption{Evolution of storm surge signal, presented as Hov-Möller diagram showing the residual water level for the stations shown in Figure \ref{fig:map_selected} for observations, NORA-Surge and Flo trained on 4 and 16 years from hindcast. The plots in panel $(a)$ cover a 7-day time period in October 2010 that had modest amplitudes and variations in residual water level, while panel $(b)$ covers a four-day period containing the extreme storm surge during Storm Xaver in December 2013. The white dotted line drawn in all four plots in both panels shows the propagation through time and space of a wave traveling at $22~\mathrm{ms^{-1}}$ through the time and position of the maximum observed residual water level. This phase speed has been chosen based on the assumption that the average depth of the southern North Sea is approximately $50~\mathrm{m}$, and that the coastal Kelvin wave propagates at the speed of a shallow water wave $(c=\sqrt{gh})$. Values of residual water level are extracted at each station, and linearly interpolated between the stations.}
    \label{fig:hovmoeller}
\end{figure}

\begin{figure}
    \centering
    \subfloat[]{\includegraphics[width=0.75\linewidth]{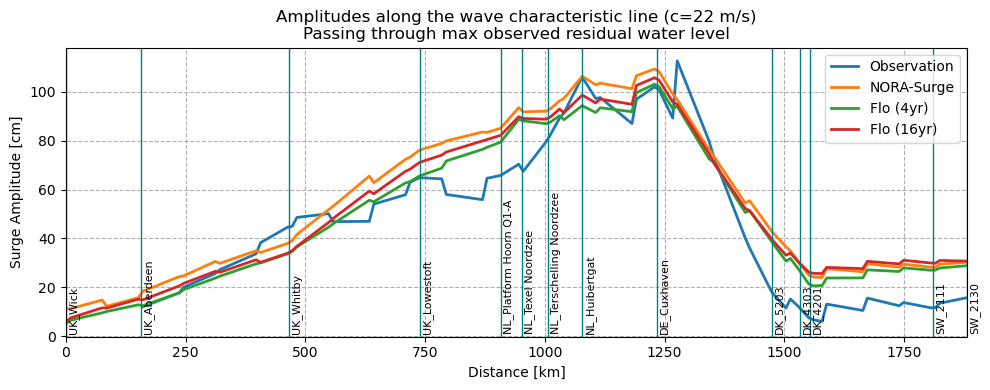}}\hfill
    \subfloat[]{\includegraphics[width=0.75\linewidth]{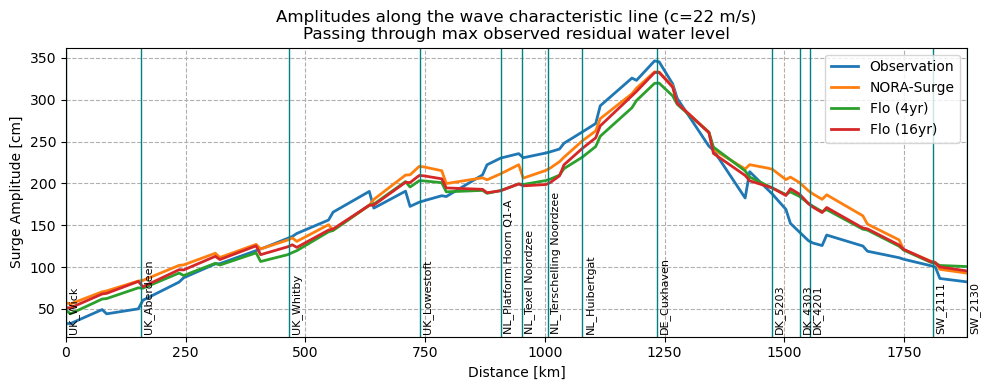}}
    \caption{Residual water level amplitudes along the white dotted line in Figure \ref{fig:hovmoeller}, representing the shallow water wave traveling at a constant phase speed of $22~\mathrm{ms^{-1}}$. The positions of the stations in Figure \ref{fig:map_selected} are plotted along the vertical lines in both panels. Panel $(a)$ and $(b)$ in this figure corresponds to panel $(a)$ and $(b)$ in Figure \ref{fig:hovmoeller}, respectively.}
    \label{fig:along_wave}
\end{figure}

A key feature of a model used for prediction of water level is its ability to resolve the propagation of a barotropic coastal Kelvin wave. Even if the Flo model does not contain explicit physics for describing the relevant processes for simulating Kelvin waves, we evaluate how well the model has learned the physics during the training. The Hov-Möller diagrams in Figure \ref{fig:hovmoeller} depict the residual water level at the stations as shown in Figure \ref{fig:map_selected}, going counter-clockwise from "UK\_Wick" (Wick, United Kingdom) to "SW\_2130" (Kungsvik, Sweden), a distance of almost $1900~\mathrm{km}$. 
We include a comparison of the Hov-Möller diagrams for both observations, NORA-Surge and Flo trained on both 4 and 16 years of training data. 

Panel $(a)$ shows a 7-day period in October 2010 with modest amplitudes and variations in the residual water level, whereas panel $(b)$ visualizes the residual water level for four days during the Storm Xaver (further discussed in Section \ref{sec:xaver}). The white dotted line drawn in all four plots in both panels shows the propagation through time and space of a wave traveling at $22~\mathrm{ms^{-1}}$ through the time and position of the maximum observed residual water level. This phase speed has been chosen based on the assumption  that the coastal Kelvin wave propagates at the speed of a shallow water wave $(c=\sqrt{gh})$, with the average depth of about $50~\mathrm{m}$ for the southern North Sea. Residual water-level values are extracted at each station and linearly interpolated between stations. 

In figure \ref{fig:along_wave} we show the amplitude of the residual water level for the observations and models along the white dotted line in Figure \ref{fig:hovmoeller}, which is interpreted as the phase-following amplitude of the Kelvin wave.
These two plots clearly show how the Flo model, based on both 4 and 16 years of training, is able to accurately simulate the correct phase speed and amplitude of a coastal Kelvin wave traveling along the coast of the entire southern North Sea.

\subsection{Comparison against observations}\label{sec:compare_obs}
One of the most important metrics of any storm surge model is how well it compares to observations.
We compare against the observations introduced in Section \ref{sec:obsdata}. Prior to comparison, we remove the mean from both the observation and model datasets to account for differences in sea surface reference levels.
The evaluation focuses on the averaged root-mean-squared-error (RMSE, see Equation \ref{eq:rmse}) and correlation, and we compare the hindcasts (see Section \ref{sec:infer}) based on both the short and long training period, and the NORA-Surge, against the observations. 

To explore how well the three hindcasts simulate large water level amplitudes
, we evaluate and validate the results after masking time periods with absolute values of the observed residual water level less than 10 and 30 cm.
Since the different stations have different distributions of residual water level, and it is common that higher absolute values of residual water level result in higher RMSE values, we also include the averaged RMSE scaled by the standard deviation of the observed residual water level for each station, denoted as $\sigma_s$ in Equation \ref{eq:rmse_s}. This yields a relative averaged RMSE based on scaled values of the local RMSE, that can be compared across different regions and stations with variable distribution of residual water level.

\begin{equation}
    \mbox{RMSE} = \sqrt{\frac{\sum_{i=1}^{n} (\mbox{obs}_i - \mbox{mod}_i)^2}{n}},
    \label{eq:rmse}
\end{equation}

\begin{equation}
    \mbox{RMSE}/\sigma = \sqrt{\frac{\sum_{i=1}^{n} (\frac{\mbox{obs}_i - \mbox{mod}_i}{\sigma_s})^2}{n}},
    \label{eq:rmse_s}
\end{equation}

The results for all stations combined are displayed in Table \ref{tab:stats_all}, and owing to the fact that our main geographical area of interest is the coast of Norway, we include the results calculated for the Norwegian stations only in Table \ref{tab:stats_nor}.

\begin{figure}
  \centering
  \subfloat[NORA-Surge]{\includegraphics[width=0.33\textwidth]{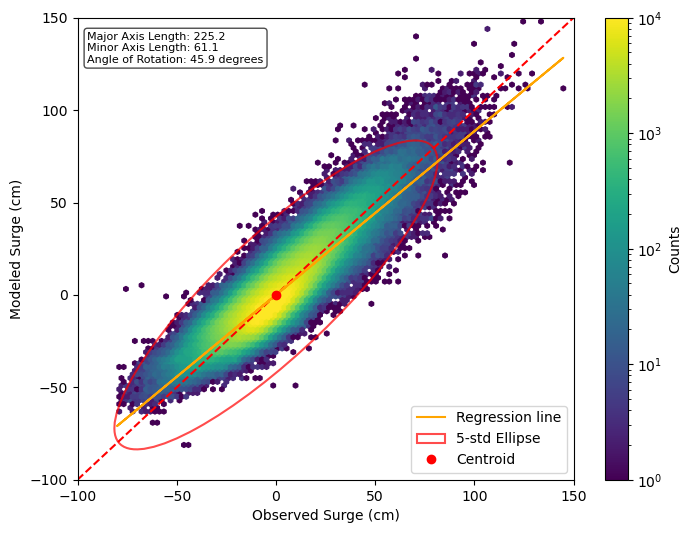}\label{fig:heat_nora_nor}}
  \subfloat[Train 16 years]{\includegraphics[width=0.33\textwidth]{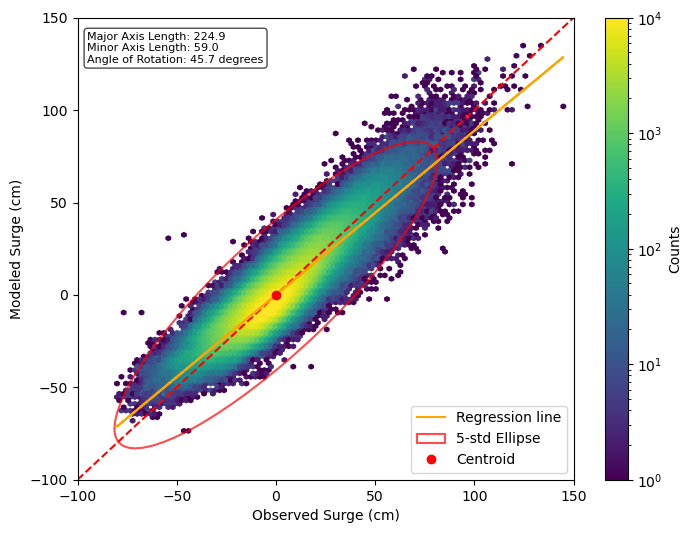}\label{fig:heat_16yr_nor}}
  \subfloat[Train 4 years]{\includegraphics[width=0.33\textwidth]{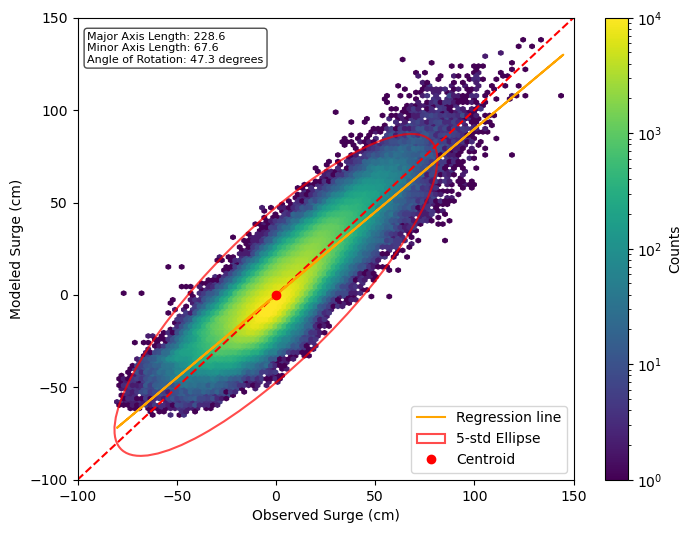}\label{fig:heat_4yr_nor}}
  \caption{Heatmap scatter plots comparing the hindcast and observations for the Norwegian stations. Note the log scale for count. The ellipsoid displayed shows the area of 5 standard deviations along, and normal to, the diagonal, centered at the point of maximum distribution. Lengths of the major and minor axes are displayed in each plot, and provide a measure of the spread around the centroid along the diagonal.}
  \label{fig:heat_scatter_nor}
\end{figure}

The focus on the Norwegian stations is continued in Figure \ref{fig:heat_scatter_nor}, where we display the heatmap scatter plots that compare the hindcast and observations for the Norwegian stations. In addition, we display an ellipsoid that shows the area of 5 standard deviations along and normal to the diagonal, centered at the point of maximum distribution. Lengths of the major and minor axes are displayed in each plot, and provide a measure of the spread around the centroid along the diagonal. 
Comparing the lengths of the minor axes of the ellipsoids for the three hindcasts, we demonstrate that the Flo model based on the 16-year training yields slightly lower error, i.e., spread around the diagonal, with the NORA-Surge and the short training at slightly higher errors. This is also reflected in the RMSE shown in Table\ref{tab:stats_nor}. 

Overall, Flo has performance and error statistics comparable to those of NORA-Surge. This is as expected, since Flo is trained on NORA-Surge, and therefore aim to emulate the behavior of NORA-Surge. In the cases where Flo has lower error than its numerical counterpart, this is likely an effect of the smoothing of the model fields, so the effect of double penalty is reduced. This is a common and well-known mathematical result when using MSE loss functions in the training of GNN-based ML models, as shown by e.g. \cite{graphcast2023} and \cite{nipen2025}.

\subsection{Case study of Storm Xaver}\label{sec:xaver}
Storm Xaver moved slowly from west to east across Northern Europe during 5-7 December 2013 \citep{deutschlander:etal:2013, kristensen2024, kristensen2025}. The strong northerly winds over the North Sea resulted initially in a convergence of water along the eastern coast of the United Kingdom due to Ekman transport of water towards the coast. When the wind direction started to back left, this water started to move anti-clockwise along the North Sea coastline as a free Kelvin wave. The combined Kelvin wave with local amplifications due to Ekman transport by the wind, and the inverse barometer effect resulted in a storm surge among the top five highest recordings over the last 100 years in the German Bight according to \citealt{deutschlander:etal:2013}. 

As the low pressure system moved further east, and the wind direction continued backing further left, the storm surge signal moved further through the North Sea up along the western coast of Denmark and into the Skagerrak area as a Kelvin wave (see \citealt{kristensen2024}). 

\begin{figure}
    \centering
    \includegraphics[width=1\linewidth]{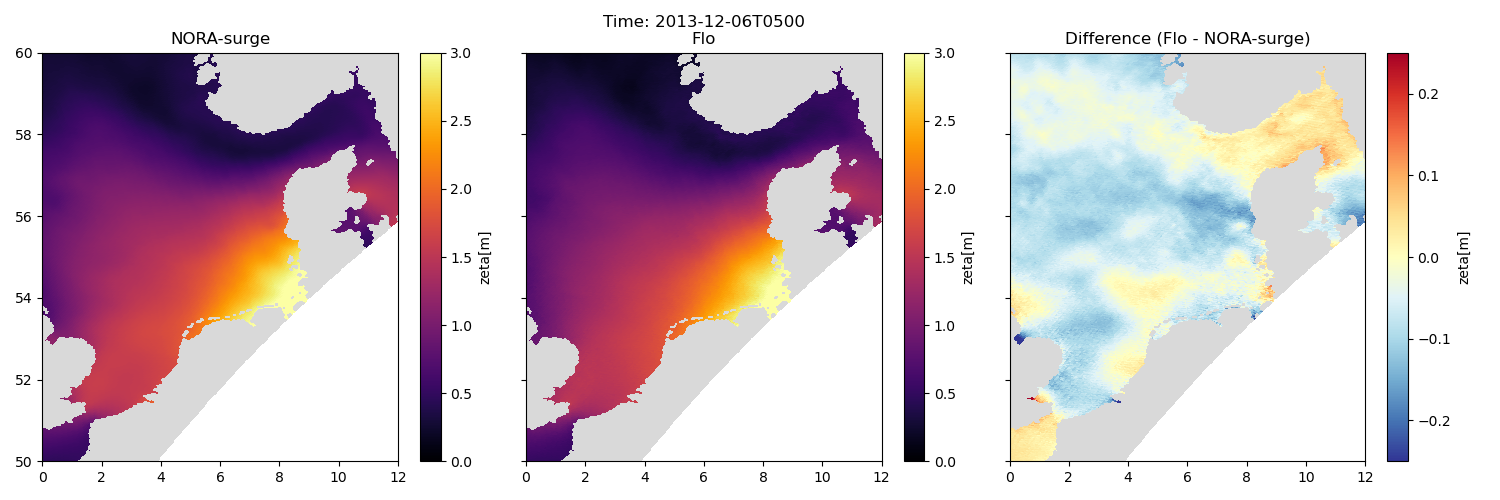}
    \caption{Comparison of residual water level between Flo and NORA-Surge at the peak of the storm Xaver in the German Bight at 05 UTC on the 6th of December 2013. Left panel is NORA-Surge, the middle panel is Flo and the right panel is the difference between the two.}
    \label{fig:xaver_compare_field}
\end{figure}

\begin{figure}
    \centering
    \includegraphics[width=0.75\linewidth]{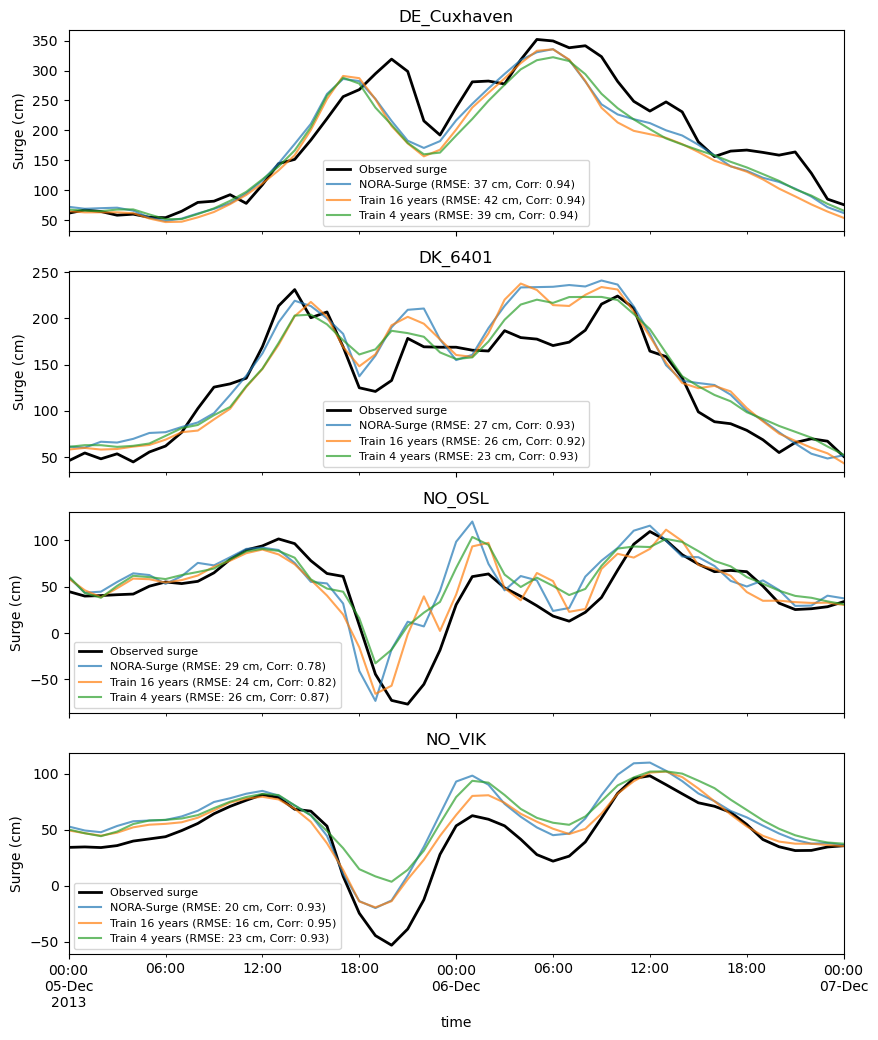}
    \caption{Time series comparison of the observed residual water level for four selected stations compared to the data-driven Flo model and NORA-Surge during the storm Xaver in December 2013. The stations are, from top to bottom, Cuxhaven (GER), Esbjerg (DEN), Oslo (NOR) and Viker (NOR). RMSE and correlation for the depicted time period for each of the model runs is given in the legend for each station.}
    \label{fig:xaver_timeseries}
\end{figure}

The storm Xaver, due to its characteristics as an extreme and special event, has been thoroughly studied by e.g. \cite{staneva17} and \cite{kristensen2024, kristensen2025}. Figure \ref{fig:xaver_compare_field} shows the residual water level in the North Sea from the Flo model compared to NORA-Surge at the peak of the storm surge in the German Bight during storm Xaver. The fields from the two hindcasts are visually very similar, and the difference-plot confirms that even the largest differences are less than 20 cm. The general impression of the difference is that the Flo hindcast has lower amplitudes than NORA-Surge in most places, with a few exceptions. An area that stands out from the rest is the Skagerrak area between Norway, Denmark and Sweden, where the Flo hindcast has higher amplitudes than NORA-Surge. This is generally an area where NORA-Surge, and the operational model based on the same model configuration, struggle (see \cite{kristensen22, tedesco23, kristensen2024}). 
The time series in Figure \ref{fig:xaver_timeseries} show a comparison between observations and the three hindcasts for four selected stations from Cuxhaven (GER) in the south, up to Oslo (NOR) in the north. This clearly shows how the Flo model, based on both the 16- and 4-year training periods, has comparable performance as the NORA-Surge for the storm Xaver, both with regards to RMSE and correlation.

This suggests that, contrary to some DDMs for the atmosphere and weather prediction, that emphasize their shortcomings when it comes to predicting extremes (e.g. \cite{nipen2025}), the Flo model is able to reproduce an extreme event like storm Xaver with similar accuracy as the NORA-Surge based on a numerical model. We note that this finding based on a single event should not be generalized, and more cases should be analyzed to verify its consistency.

\section{Summary and Conclusions}\label{sec:summary}
The data-driven storm surge model \textit{Flo} is presented, capable of simulating reliable residual water level predictions for the limited area model domain covering the North Sea, the Norwegian Sea and the Barents Sea. Flo produces water level simulations at a horizontal resolution of 4 km, and a temporal resolution of 1 hour. The model is built using the Anemoi framework \citep{lang2024}, and uses a GNN to predict the residual water level as a continuous two-dimensional field for any arbitrary location within the model domain. 
This sets Flo apart from most known previous efforts to create storm surge models using ML where architectures like e.g. ANNs, CNNs, and LSTMs were used to provide predictions only at pre-described locations.
The training of Flo is based on the NORA3 hindcast \citep{haakenstad21nora3, haakenstad22} for atmospheric forcing (MSLP and 10-meter winds), and the NORA-Surge hindcast \citep{kristensen2024} for the residual water level. Both hindcasts cover the 43-year period from 1979 to 2022, and we utilize the 20-year period from 1990 to 2009 for training and validation. The years 2010-2022 were reserved for testing and evaluation.

The hindcasts covering the period 2010 to 2018 created by Flo have been qualitatively evaluated and validated against both the NORA-Surge hindcast and a dataset of observations of residual water level from 98 water level stations. The evaluation demonstrates that Flo has a comparable performance with regard to RMSE and correlation as its numerical counterpart, here represented by the NORA-Surge hindcast. When averaging over the entire time period of the Flo hindcast, and all stations, NORA-Surge yields an RMSE of 12.6 cm, while Flo yields an RSME of 12.4 cm. This very slight improvement holds even when masking out the time periods with observed absolute values of residual water level of less than 10 and 30 cm. We note that this slightly lower RMSE value is a likely effect of smoothing of the model fields, which is a well-known mathematical result when using the MSE loss function in the training of the GNN model. This smoothing inherently reduces the effect of double penalty in the RMSE.

The case study of the extreme winter storm Xaver of December 2013, which resulted in one of the highest water levels in the German Bight in the last 100 years, shows how Flo is capable of predicting an extreme and rare storm surge event with a quality similar to that of NORA-Surge, even if the limited training data set contains very few data points of similar magnitude. Moreover, the ability to physically correctly predict the amplitude and propagation of the resulting coastal Kelvin-wave shows how Flo is able to learn, in addition to the direct effect of MSLP gradients and wind stress, the more complex underlying physics that drives important parts of the storm surge signal.

We acknowledge the fact that developing and training a DDM, like Flo, is a costly process with regards to computing resources, and is critically dependent on having an existing hindcast or reanalysis for training. With this in mind, a relevant question is whether the marginal improvements that this model currently realizes can be justified. We anticipate that the full potential of transitioning from a purely numerical model to a data-driven machine learning model lies not in the work presented here, but in the emerging potential for future developments. As a next step, observations could be included in the model training to improve the predictions through optimal interpolations, multi-encoder architecture or physics embedded neural networks. By doing this, the DDM may achieve more accurate predictions than the numerical model, since it is not limited by the discretizations of the governing equations of motion, but is able to find statistical relations in the training data, which includes information from observations. In our case, instead of running a costly reanalysis using a numerical model, the inclusion of information from observations may be performed to overcome model biases in the hindcast.

Future development will, in addition to the inclusion of observation data to increase model skill beyond its numerical counterpart, involve further training on operational forecast data from the numerical model used for storm surge prediction \citep{kristensen22}. The number of training steps and how many years are used in training should be explored further. Additionally, the use of ensemble model data in training can provide probabilistic predictions by the Flo model in an operational value chain, and its use in decision-making during impact-based forecasting services.

\subsection*{Open research section}
The Anemoi framework is freely available on Github: \href{https://github.com/ecmwf/anemoi-core}{\url{https://github.com/ecmwf/anemoi-core}}. All the config files used in the Flo training and inference are also available on Github: \href{https://github.com/nilsmkMET/anemoi-stormsurge/tree/main}{\url{https://github.com/nilsmkMET/anemoi-stormsurge/tree/main}}. The NORA-Surge hindcast can be downloaded as netCDF-files from the MET Norway Thredds server: \href{https://thredds.met.no/thredds/catalog/stormrisk/catalog.html}{\url{https://thredds.met.no/thredds/catalog/stormrisk/catalog.html}}. And the NORA3 atmospheric hindcast data can be accessed here: \href{https://thredds.met.no/thredds/catalog/nora3_subset_atmos/catalog.html}{\url{https://thredds.met.no/thredds/catalog/nora3_subset_atmos/catalog.html}}. Finally, the ERA5 dataset used to fill the atmospheric fields outside the coverage of the NORA3 domain, is available through the Copernicus Climate Data Store: \href{https://cds.climate.copernicus.eu/datasets/reanalysis-era5-single-levels}{\url{https://cds.climate.copernicus.eu/datasets/reanalysis-era5-single-levels}}.

\clearpage
\begin{table}
    \centering
    \begin{tabular}{lll} \hline
         \textbf{Variable} & \textbf{Source}  & \textbf{Type} \\ \hline
         SSH (residual water level) & NORA-Surge & Prognostic \\
         10 meter wind (U10m and V10m) & NORA3 & Forcing \\
         MSLP & NORA3 & Forcing \\
         Bathymetry & NORA-Surge & Forcing (static) \\
         Land-sea mask & NORA-Surge & Forcing (static) \\
         Coriolis parameter & NORA-Surge & Forcing (static) \\
         Solar insolation & Anemoi & Forcing (static) \\
         Sine of latitude & Anemoi & Forcing (static) \\
         Sine of longitude & Anemoi & Forcing (static) \\
         Sine of local time & Anemoi & Forcing (static) \\
         Sine of Julian day & Anemoi & Forcing (static) \\
         Cosine of latitude & Anemoi & Forcing (static) \\
         Cosine of longitude & Anemoi & Forcing (static) \\
         Cosine of local time & Anemoi & Forcing (static) \\
         Cosine of Julian day & Anemoi & Forcing (static) \\ \hline
    \end{tabular}
    \caption{List of the variables used in the training of Flo, including their source and type (i.e. if they are forcing or prognostic).}
    \label{tab:trainvars}
\end{table}

\begin{table}
    \centering
    \begin{tabular}{|ll|c|c|c|}
        \hline
        \textbf{Exp. name} &  & \textbf{RMSE [cm]} & \textbf{RMSE$/\sigma$ [$\sigma$]} & \textbf{Correlation} \\ \hline
        NORA-Surge         & All data             & 12.6 & 0.55 & 0.85 \\
                           & Masked $< \pm 10 cm$ & 13.4 & 0.58 & 0.90 \\
                           & Masked $< \pm 30 cm$ & 16.7 & 0.71 & 0.94 \\ \hline
        Train 16 years     & All data             & 12.4 & 0.54 & 0.86 \\
                           & Masked $< \pm 10 cm$ & 13.2 & 0.57 & 0.90 \\
                           & Masked $< \pm 30 cm$ & 16.6 & 0.70 & 0.94 \\ \hline
        Train 4 years      & All data             & 13.1 & 0.58 & 0.84 \\
                           & Masked $< \pm 10 cm$ & 13.9 & 0.61 & 0.89 \\
                           & Masked $< \pm 30 cm$ & 17.4 & 0.75 & 0.94 \\ \hline
    \end{tabular}
    \caption{Averaged statistics for all 98 stations over the time period 2010-2018. The rows labeled "All data" contain all valid data from the comparison against observations. The rows labeled "Masked" contain only the data for when the absolute values of the observed residual water level is larger than the given value. I.e. all data for when the absolute value of the observations is less is masked out. This is done to explore how well the different models handle larger values of residual water level (extremes).}
    \label{tab:stats_all}
\end{table}

\begin{table}
    \centering
    \begin{tabular}{|ll|c|c|c|}
        \hline
        \textbf{Exp. name} & & \textbf{RMSE [cm]} & \textbf{RMSE$/\sigma$ [$\sigma$]} & \textbf{Correlation} \\ \hline
        NORA-Surge         & All data             & 8.4 & 0.53 & 0.87 \\
                           & Masked $< \pm 10 cm$ & 8.8 & 0.56 & 0.92 \\
                           & Masked $< \pm 30 cm$ & 10.3 & 0.65 & 0.97 \\ \hline
        Train 16 years     & All data             & 8.1 & 0.51 & 0.88 \\
                           & Masked $< \pm 10 cm$ & 8.5 & 0.54 & 0.93 \\
                           & Masked $< \pm 30 cm$ & 10.0 & 0.63 & 0.97 \\ \hline
        Train 4 years      & All data             & 9.2 & 0.58 & 0.85 \\
                           & Masked $< \pm 10 cm$ & 9.6 & 0.61 & 0.91 \\
                           & Masked $< \pm 30 cm$ & 11.2 & 0.71 & 0.97 \\ \hline
    \end{tabular}
    \caption{Same as Table \ref{tab:stats_all}, but for the 24 Norwegian stations only.}
    \label{tab:stats_nor}
\end{table}

\bibliographystyle{cas-model2-names}

\bibliography{surge}

\end{document}